\RequirePackage{lineno}
\documentclass[aps,prb,preprint,superscriptaddress]{revtex4}

\usepackage{graphicx}
\usepackage{bm}
\usepackage{amsmath}
\usepackage{amssymb}
\usepackage{hyperref}
\usepackage[version=3]{mhchem}
\usepackage{float}

\begin{document}

\title{Role of force-constant difference in phonon scattering by nano-precipitates in PbTe}

\author{Xiaolong Yang}
\affiliation{Frontier Institute of Science and Technology, and State Key Laboratory for Mechanical Behavior of Materials, Xi'an Jiaotong University, 710054, Xi'an, P. R. China.}

\author{Jes\'us Carrete}
\affiliation{LITEN, CEA-Grenoble, 17 rue des Martyrs, 38054 Grenoble Cedex 9, France}

\author{Zhao Wang}
\email{wzzhao@yahoo.fr}
\affiliation{Frontier Institute of Science and Technology, and State Key Laboratory for Mechanical Behavior of Materials, Xi'an Jiaotong University, 710054, Xi'an, P. R. China.}

\begin{abstract}
We study the effect of nanoscale precipitates on lattice thermal conduction in thermoelectric PbTe using a combination of \textit{ab-initio} phonon calculations and molecular dynamics. We take into account the effects of mass differences and changes in force constants, and find an enhanced influence of the latter with increased precipitate concentration. As a consequence, our inclusion of the changes in force constants in the calculation affords a smaller predicted optimal nano-precipitate size that minimizes the thermal conductivity. These results suggest that the phonon scattering by nanoprecipitates in thermoelectric composites could be stronger than previously thought.
\end{abstract}

\maketitle

\section{Introduction}

Thermoelectric materials, capable of converting heat into electric power and vice versa, have attracted increasing interest for applications in energy harvesting and interconnection technologies because of their solid-state features: no moving parts, quiet operation, high reliability and so forth.\cite{Zebarjadi2012,Nielsch2011,Minnich2009} The efficiency of a thermoelectric material at a temperature $T$ can be gauged by its dimensionless figure of merit $ZT = (S^2\sigma/\kappa)T$, where $S$, $\sigma$ and $\kappa$ are its Seebeck coefficient, electrical conductivity and thermal conductivity, respectively. Nanostructuring is considered as a crucial strategy to enhance $ZT$ by lowering the thermal conductivity by scattering phonons at nanoscale interfaces and defects.\cite{Androulakis2007,Hsu2004,Biswas2011} A typical approach involves embedding nanometer-sized precipitates in bulk thermoelectric composites.\cite{Pei2011,Kim2006,Hsu2004,Biswas2011,Ikeda2012,He2010} To this end, many efforts have been devoted into the development of synthesis methods to embed nano-particles into bulk thermoelectrics using ball milling, continuous gas-phase synthesis and subsequent compacting based on hot-pressing or spark plasma sintering.\cite{Tang2007,Ma2008,Cao2008,Xie2009,Ebling2007,Yan2010,Pei2011} High $ZT$ values of about $1.4$ at $373\,\mathrm{K}$ for BiSbTe bulk alloys with Te nanoprecipitates,\cite{Poudel2008} and about $1.47$ at $700\,\mathrm{K}$ for BiSbTe with BiTe and SbTe nanoprecipitates,\cite{Cao2008} have been reported using these methods.

The physical role of nanoparticles in phonon scattering was studied by Kim and Majumdar,\cite{Kim2006a} and by Mingo and co-workers,\cite{Mingo2009} and by Wu and co-workers\cite{Wu2013} based on earlier works of Klemens \textit{et al.}.\cite{KLEMENS1955,TURK1974} Note that a very-recently work showed that strongly concentrated, bimodal particle size distributions could lower the lattice thermal conductivity of SiGe beyond the single-size limit.\cite{Zhang2015} In this kind of setting, two mechanisms contribute to phonon scattering: the mass difference between the filler and the matrix alloy, and the differences in the interatomic force constants due to changes in chemical bonding. The application of those models seems to be successful in interpreting a wide range of experimentally measured data.\cite{He2010,He2012,Lo2012,He2012a} It must be noted, however, that the contribution of the force-constant differences is often neglected\cite{Mingo2009} due to the lack of atomistic-scale information despite its possible importance.\cite{Murakami2013,Kim2006a,katcho_effect_2014}

Here we report a quantitative analysis of the mechanisms of phonon scattering by Pb embedded in a PbTe alloy in order to highlight the importance of this often-neglected part. We combine atomistic simulations with the nanoparticle-scattering-limited phonon relaxation time approach. Thermoelectric PbTe was chosen in this work as a typical example of a high-performance thermoelectric material \cite{Yang2013} with nanometer-scale precipitates frequently observed in experiments.\cite{He2010,Wu2014,He2012,Lo2012,He2013,He2012a}

\section{Methods}
\subsection{Molecular dynamics}

In our molecular dynamics simulations, atomistic interactions are described by a two-body Buckingham potential, which has been shown to successfully reproduce the mechanical and phonon properties of bulk PbTe.\cite{Qiu2012,Qiu2012a} Detailed parameters for the potential are given in Ref. \onlinecite{Qiu2012}. Electrostatic interactions are taken into account by means of Ewald summation with an appropriate choice of parameters.\cite{Wolf1999} We use the parallel molecular dynamics package LAMMPS\cite{PLIMPTON1995} with a velocity Verlet algorithm for numerical integration of the equations of motion in a time step $dt=0.5\,\mathrm{fs}$.\cite{Yang2015,Li2013}

\begin{figure}[thp]
\centerline{\includegraphics[width=11cm]{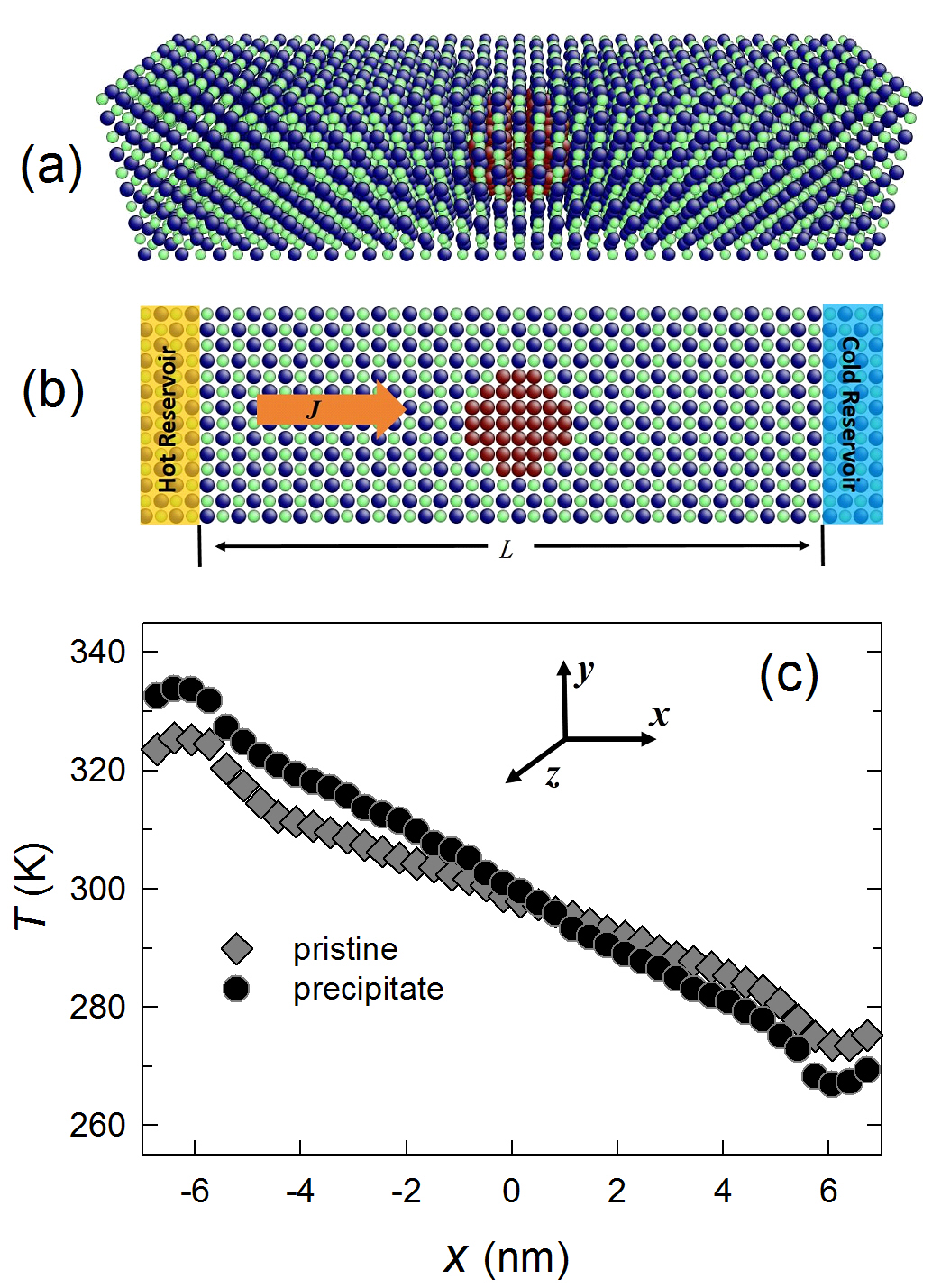}}
\caption{\label{fig1}
(a) Snapshot of the simulation cell. Blue, green and red atoms represent Pb at PbTe crystal sites, Te, and Pb in a precipitate, respectively. (b) Schematic description of non-equilibrium molecular dynamics simulation. (c) Temperature profile along the longitudinal (\textit{x}) direction in two different samples without and with a Pb nano-precipitate.}
\end{figure}

An example of our simulation cell is shown in Fig. \ref{fig1}(a). In order to build Pb nano-precipitates consistent with those observed in experiments,\cite{He2010,Wu2014,He2012,Lo2012,He2013,He2012a} we insert a spherical Pb cluster with a diameter of about $2\,\mathrm{nm}$ (red circles) into a pristine PbTe crystal. Fig. \ref{fig1}(b) shows a schematic diagram of our non-equilibrium molecular dynamics (using more specifically,\cite{schelling} the so-called direct method) simulation. The simulation cell consists of a sample region of length $L$ between a hot and a cold reservoirs. Periodic boundary conditions are applied in the transverse (\textit{y} and \textit{z}) directions, and a heat flux $J$ is enforced along $x$. We wait for the system to reach the stationary state and start recording the temperature gradient $\partial T/\partial x$ during the simulation, as illustrated in Fig. \ref{fig1}(c). The lattice thermal conductivity $\kappa$ is then extracted from the Fourier's law,

\begin{equation}
\label{eq:1}
\kappa=-\frac{J}{\partial T/\partial x}.
\end{equation}

More specifically, the system is first relaxed in the isothermal-isobaric (NPT) ensemble using the Nos\'{e}-Hoover thermostat for  $5 \cdot 10^{6}$ time steps. We then switch to the canonical (NVE) ensemble for a further $10^8$ time steps to gather the required statistics.

\subsection{Boltzmann transport framework and \textit{ab-initio} phonon calculation}

Under the relaxation-time approximation (RTA),\cite{Wang2010,Wang2011} the lattice thermal conductivity can be written as

\begin{equation}
\label{eq:2}
\kappa (T)=\frac{1}{3}\sum_{i}\int\frac{d^3\mathbf{q}}{8\pi^3}\upsilon_{iq}^2\tau_{iq}c_{iq},
\end{equation}

\noindent where the sum runs over all phonon bands, the integral extends over the whole first Brillouin zone, $\upsilon_{iq}$ is the group velocity of a given phonon mode, $\tau_{iq}$ is its relaxation time, and $c_{iq}$ its contribution to the volumetric heat capacity. The total relaxation time can be approximated by a  Matthiessen sum of the anharmonic ($\tau_{a}$), boundary ($\tau_{b}$), and nanoparticle ($\tau_{np}$) contributions,

\begin{equation}
\label{eq:3}
\tau^{-1}=\tau_{a}^{-1}+\tau_{b}^{-1}+\tau_{np}^{-1}.
\end{equation}

The three-phonon anharmonic term $\tau_{a}$ is calculated using the model proposed by Slack \textit{et al.}\cite{SLACK1964,Morelli2002}

\begin{equation}
\label{eq:4}
\tau_{a}^{-1}=p \omega^{2}\frac{T}{\theta_{D}}e^{\frac{-\theta_{D}}{3T}}.
\end{equation}

\noindent Here, the $p$ parameter is fitted to the bulk thermal conductivity of pristine PbTe computed by molecular dynamics. The Debye temperature ($\theta_{D}$) is obtained from the second moment of the distribution of phonon frequencies:\cite{Bjerg2014}

\begin{equation}
\label{eq:5}
\theta_{D}=n^{-\frac{1}{3}}\sqrt{\frac{5\hbar^2}{3k_{B}^2}\frac{\int_{0}^{\infty}\omega^2 g(\omega)d\omega}{\int_{0}^{\infty}g(\omega)d\omega}},
\end{equation}

\noindent where $n$ is the number of atoms per unit cell, $g(\omega)$ denotes the phonon density of states, $k_{B}$ is the Boltzmann constant.

The boundary scattering term\cite{Wang2011a} is estimated as the group velocity divided by the simulation box length $L$ times a form factor $F$.\cite{Takashiri2012}

\begin{equation}
\label{eq:6}
\tau_{b}^{-1}=F\frac{\upsilon}{L}.
\end{equation}

\noindent Like $p$ above, $F$ is fitted to molecular dynamics data using the least-squares method at a given initial value.

The nanoprecipitate term $\tau_{np}$ is obtained by using a Mathiessen interpolation between the long (Rayleigh) and short (geometric) wavelength scattering regimes:\cite{Kim2006a,JOSHI1993}

\begin{equation}
\label{eq:7}
\tau_{np}^{-1}=\upsilon_{g}(\sigma_{s}^{-1}+\sigma_{l}^{-1})^{-1}V_{\rho},
\end{equation}

\noindent where $V_{\rho}$ is the number density of nanoparticles, $\upsilon_{g}$ is the phonon group velocity, $\sigma_s$ is the scattering cross section in the short-wavelength limit, and $\sigma_l$ is the cross section in the long-wavelength limit,

\begin{align}
  \sigma_{s}&=\frac{\pi D^2}{2},\label{eq:8}\\
  \sigma_{l}&=\frac{1}{9}\pi D^{2}[\left(\Delta \rho/\rho\right)^{2}+12\left(\Delta K/K\right)^{2}]\left(\frac{\omega D}{2 \upsilon_{g}}\right)^{4},\label{eq:9}
\end{align}

\noindent where $D$ is the particle diameter, $\rho$ is the mass density of the medium  and $\Delta \rho$ is the mass density difference between the particle and matrix materials, $K$ is the host force constant and  $\Delta K$ is the force-constants difference between the particle and matrix materials, where they are replaced by elastic constants.\cite{Petersen2014,Jong2015}

\begin{table}[tbp]
\centering
\caption{Input parameters for calculating the phonon scattering relaxation times of PbTe with a 1nm-sized Pb precipitate.}
\begin{tabular}{llcc}
\hline
Relaxation time  &Parameter &Symbol (unit) &Value\\ \hline
$\tau_{a}^{-1}$ & Fitted parameter of anharmonic scattering  & $p\,(\mathrm{ps})$ &0.0040376\\
		 &Debye temperature of PbTe  & $\theta_{D}\,(\mathrm{K})$ &136\\
		&Volume per atom in PbTe & $V\,(\mathrm{nm}^{3})$ &0.03369\\  \hline
$\tau_{b}^{-1}$ &Grain size & $L\,(\mathrm{nm})$ &10\\
				&Fitted form factor &$F$ &0.239326\\ \hline
$\tau_{np}^{-1}$  &Average diameter of nanoscale precipitates & $D\,(\mathrm{nm})$ &1.0\\
		&Average sound velocity of PbTe & $\upsilon_{p}(\mathrm{m/s})$ &1776\\
		&Mass of a Pb atom &$M_{\mathrm{Pb}}\,(\mathrm{amu})$ &207.2\\
		&Mass of a Te atom &$M_{\mathrm{Te}}\,(\mathrm{amu})$ &127.6\\
		&force-constant difference &$\Delta K/K$ &1.514\\
		&Mass density difference &$\Delta \rho/\rho$ &0.389\\
		&Number density of nanoscale precipitates &$V_{\rho}\,(\mathrm{nm^{-3}})$ &0.02837\\ \hline				
\end{tabular}
\end{table}

We obtain the phonon spectrum of PbTe using a supercell-based method as implemented in the Phonopy package,\cite{Togo2008} with the  Vienna Ab initio Simulation Package (VASP)\cite{BLOCHL1994,VASP} as the density functional theory backend for computing energies and forces. We start by computing the lattice parameter that minimizes the energy of the crystal by means of a single-unit-cell calculation on a 16x16x16 Monkhorst-Pack k-point. Our result, $0.6546\,\mathrm{nm}$, is in good agreement with the experimental value of $0.6462\,\mathrm{nm}$.\cite{Kastbjerg2013,DALVEN1969} We then obtain the forces on a minimal set of displaced configurations of a $4\times 4\times 4$ supercell, as well as the high-frequency-limit dielectric tensor and a set of Born effective charges. With these ingredients we compute the harmonic interatomic force constants of the system and its phonon dispersions on a 32x32x32 q-point grid. The integral in Eq. \eqref{eq:2} is then calculated using a histogram method and its convergence evaluated by comparison with a coarser grid.

By way of example, Table I lists all the values of the parameters used in this work and involved in the thermal conductivity calculation for PbTe with a Pb precipitate with a size of $1.0\,\mathrm{nm}$ after checked the significant digits with Refs.\cite{He2010,Lo2012}

\section{Results and discussion}
\begin{figure}[thp]
\centerline{\includegraphics[width=13cm]{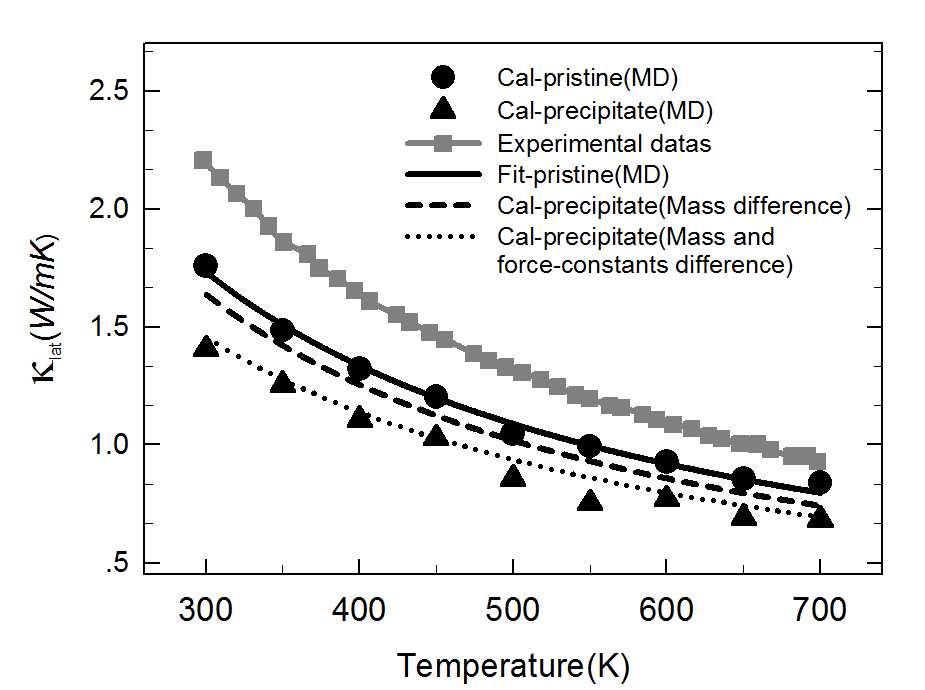}}
\caption{\label{fig2}
Temperature-dependent lattice thermal conductivity of a pristine PbTe sample and a PbTe sample with a Pb precipitate of 1 nm diameter. The symbols stand for simulated data in comparison with experimental data.\cite{Fedorov1969} The curves stand for numerical fits using the relaxation-time approximation based on our \textit{ab-initio} phonon dispersion. The dashed line is obtained taking into account only the mass difference (first term in Eq.\eqref{eq:9}), while the solid line includes the effect of the force-constant difference as well.}
\end{figure}

Using molecular dynamics simulations, we compute the thermal conductivity of both pristine PbTe and PbTe with Pb precipitates at increasing temperatures. The results are represented by the symbols in Fig. \ref{fig2}. It can be seen that our $\kappa$ values for pristine PbTe are lower than experiment.\cite{Fedorov1969} Possible causes of this discrepancy include the choice of interatomic potential or the nanometer-sized boundaries of the samples.\cite{Sellan2010}

We use the molecular-dynamics results for pristine PbTe to fit the parameters in Eqs.\eqref{eq:4} and \eqref{eq:6}, which are then applied to the numerical fit to the MD-simulated $\kappa$ of the PbTe sample containing Pb precipitates. We note that the main difference between ``Mass only'' and ``Mass and force constants'' fits is that in the latter case we use the complete form of Eqs.\eqref{eq:9}, whereas in the former we only include the first term, as it is often done in previous works, e.g. Refs.\cite{He2010,He2012,Lo2012,He2012a} As Fig. \ref{fig2} shows, the thermal conductivity is overestimated if the force-constant difference is not included.

\begin{figure}[thp]
\centerline{\includegraphics[width=11cm]{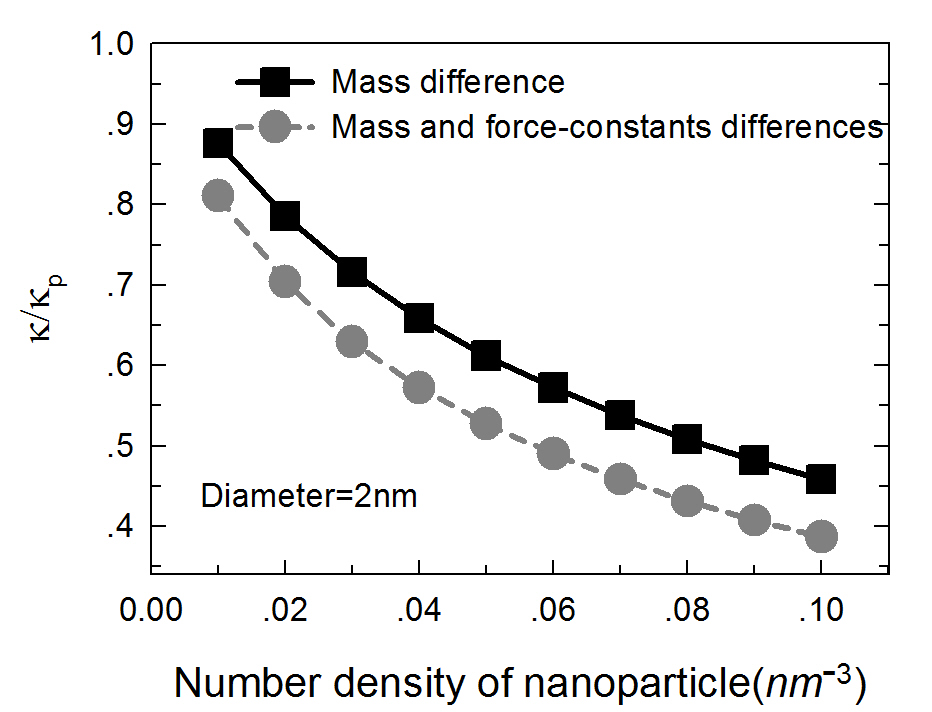}}
\caption{\label{fig3}
$\kappa/\kappa_{p}$ \textit{vs.} the number density $\rho_{n}$ of nano-precipitates with a given nanoparticle diameter $D$ of $2\,\mathrm{nm}$. Lattice thermal conductivities of PbTe nano-composites with Pb nano-precipitates $\kappa$ are normalized by those of the pristine PbTe crystal $\kappa_{p}$.}
\end{figure}

To see in what situations does the contribution from the force-constant difference become important, we compute the room-temperature thermal conductivity ratio $\kappa/\kappa_{p}$ of PbTe samples containing Pb precipitates with increasing number density $\rho_{n}$ [Fig. \ref{fig3}], using the sample numerical procedures as those for obtaining the curves shown in Fig. \ref{fig2}. It can be seen that the contribution from the force-constant difference becomes more important when the number density of  nano-precipitate increases and the contribution to phonon scattering from precipitates becomes more relevant. The reason is that nanoprecipitate phonon scattering gets gradually enhanced with increasing number density, its contribution to the total phonon scattering therefore becomes more important.

\begin{figure}[thp]
\centerline{\includegraphics[width=11cm]{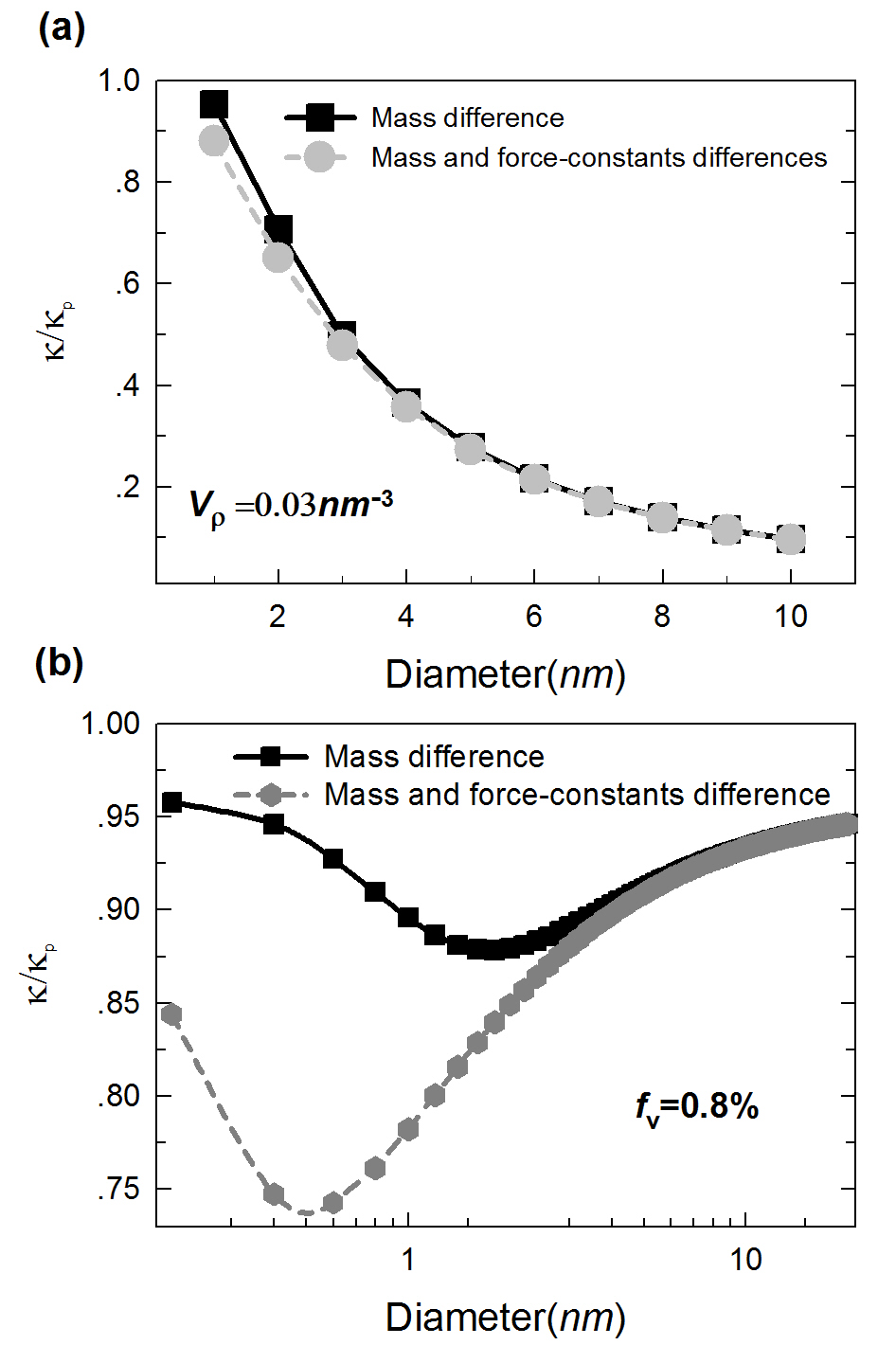}}
\caption{\label{fig4}
Lattice thermal conductivities of PbTe nano-composites with Pb nano-precipitates $\kappa$ normalized by that of the pristine PbTe crystal $\kappa_{p}$. (a) $\kappa/\kappa_{p}$ as a function of nano-precipitate size for a given number density $\rho_{n}$ of nanoparticles at $300\,\mathrm{K}$. (b) $\kappa/\kappa_{p}$ \textit{vs.} nano-precipitate size for a fixed volume fraction $\rho_{v}$ of nanoparticles at $300\,\mathrm{K}$.}
\end{figure}

In Fig. \ref{fig4} (a) it can also be seen how the force-constant difference becomes less important for large nano-precipitates. This second phenomenon can be understood by looking at Eq. \eqref{eq:8}, which interpolates between the short- and long-wavelength limits. With increasing size ($D$), the contribution of the short-wavelength term becomes more important. To illustrate a real experimental condition, we plot $\kappa/\kappa_{p}$ as a function of $D$ for a given nano-precipitate volume fraction in Fig. \ref{fig4}(b). Note that the inclusion of force-constant differences shifts the optimal size to smaller values since the estimate of phonon scattering by small-size nano-precipitates becomes higher.

\begin{figure}[thp]
\centerline{\includegraphics[width=11cm]{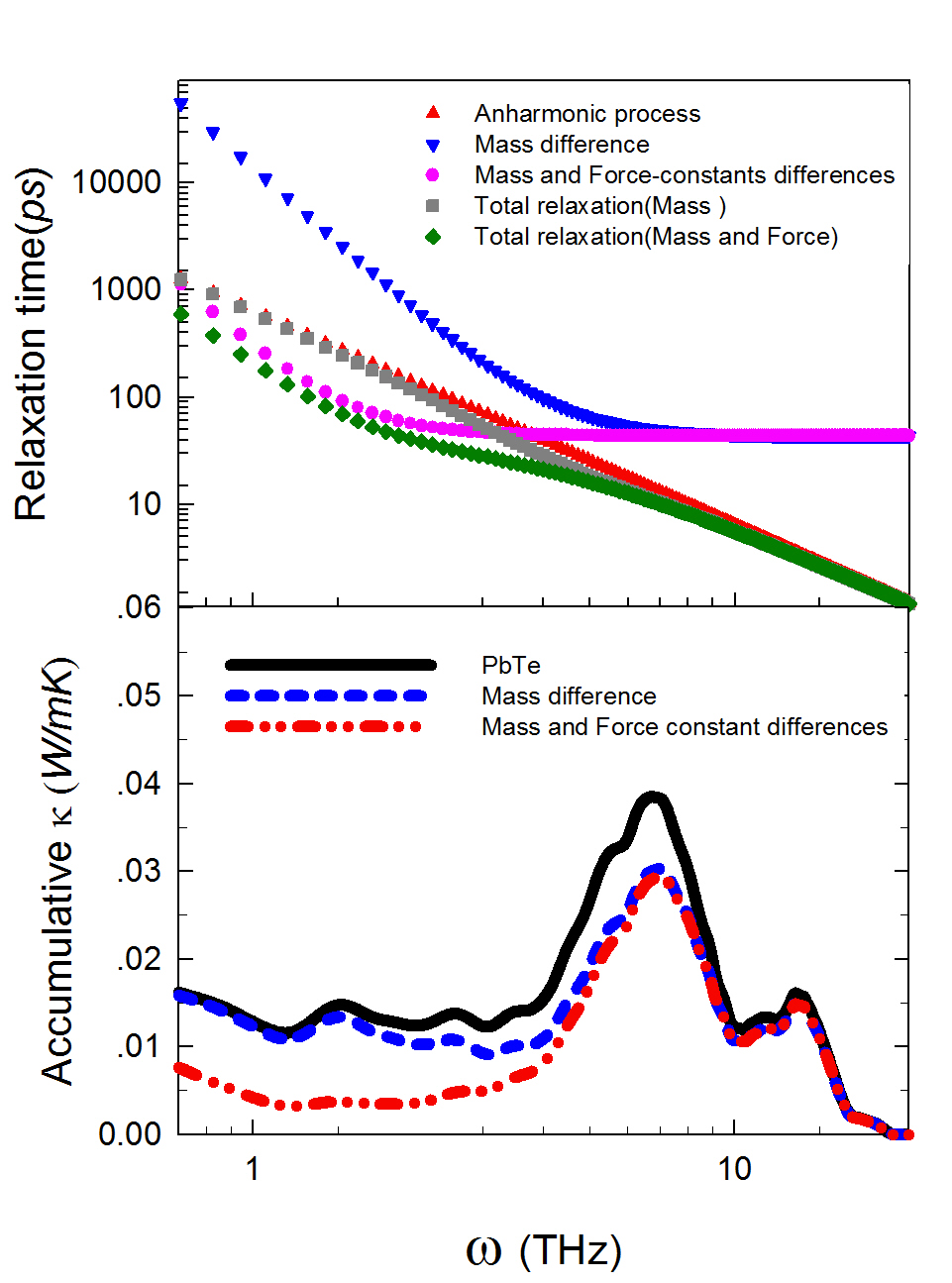}}
\caption{\label{fig5}
(a) Phonon relaxation times $\tau$ of PbTe with Pb nano-precipitates from different scattering mechanisms as a function of phonon frequency $\omega$ at $300\,\mathrm{K}$. (b) Contribution to the total lattice thermal conductivity $\kappa_a$ from each frequency at $300\,\mathrm{K}$.}
\end{figure}

Finally, using ab-initio calculation, we computed the contribution to the total relaxation time from the different scattering mechanisms considered in Eq. \eqref{eq:3} as a function of the phonon frequency $\omega$ [Fig. \ref{fig5}(a)]. It can be seen that precipitate scattering mainly influences the low-and medium-frequency phonons from $\sim 1$ to $\sim 10\,\mathrm{rad/ps}$. This frequency range makes a major contribution to the thermal conductivity [Fig. \ref{fig5}(b)]. We can also see the mass difference mainly affects the medium-frequency phonons ($~5.0-10.0\,\mathrm{rad/ps}$), while scattering coming from the force-constant difference mainly influences the low-frequency phonons ($<5.0\,\mathrm{rad/ps}$). This is also confirmed by the plot of the contribution to  the thermal conductivity from each frequency [Fig. \ref{fig5}(b)].

\section{Summary}
In summary, we computed the thermal conductivity of PbTe composites with nanometer-sized Pb precipitates using non-equilibrium molecular dynamics and the relaxation time approximation to the Boltzmann transport equation, including both the mass and force-constant differences. We find that the contribution to phonon scattering by the force-constant difference becomes important when the nano-precipitate number density increases. This contribution however diminishes with increasing precipitate size at a given number density. Fixing the volume concentration of the nanoprecipitates, we find that the optimal size that minimizes the composite thermal conductivity is reduced when both the mass and force-constant differences are considered. Detailed phonon analysis shows that the thermal conductivity reduction by nanoprecipitates originates in the enhanced phonon-phonon scattering in the low and medium-frequency ranges. These results suggest that previous approximations \cite{Mingo2009,Wu2013,He2010,He2012,Lo2012,He2012a} could overestimate the mass-difference phonon scattering by nanoprecipitates in thermoelectric composites.

\section*{Acknowledgments}
We thank  Dr. N. Mingo at C.E.A. Grenoble and Prof. J. Li at M.I.T. for helpful discussions. This work is supported by a grant-in-aid of 985 Project from Xi'an Jiaotong University, the National Natural Science Foundation of China (Grant No. 11204228), the National Basic Research Program of China (2012CB619402 and 2014CB644003) and the Fundamental Research Funds for the Central Universities.

\end{document}